\documentstyle[prb,twocolumn,aps,psfig,amssymb,floats]{revtex}

\newcommand{\ff}{f_{\sigma}^{\,\phantom{\dagger}}}
\newcommand{\ffd}{f_{\sigma}^{\,\dagger}}
\newcommand{\ffi}{f_{i\sigma}^{\,\phantom{\dagger}}}
\newcommand{\ffid}{f_{i\sigma}^{\,\dagger}}

\newcommand{\ffk}{f_{\vec{k}\sigma}^{\,\phantom{\dagger}}}
\newcommand{\ffkd}{f_{\vec{k}\sigma}^{\,\dagger}}

\newcommand{\nup}{n_{\uparrow}^{\,f}}
\newcommand{\ndown}{n_{\downarrow}^{\,f}}
\newcommand{\nsig}{n_{\sigma}^{\,f}}
\newcommand{\niup}{n_{i\uparrow}^{\,f}}
\newcommand{\nidown}{n_{i\downarrow}^{\,f}}

\newcommand{\cc}{c_{\sigma}^{\,\phantom{\dagger}}}

\newcommand{\cck}{c_{\vec{k}\sigma}^{\,\phantom{\dagger}}}
\newcommand{\cckd}{c_{\vec{k}\sigma}^{\,\dagger}}
\newcommand{\ncsig}{n_{\sigma}^{\,c}}

\newcommand{\greew}[2]{\langle\langle\,#1\, ; \,#2 \,
\rangle\rangle \! {\atop\omega}}

\newcommand{\hubb}[2]{\left| #1 \,\left\rangle \right\langle #2 \right|}

\newcommand{\del}[1]{\delta(#1)}

\newcommand{\dis}{\epsilon_{\vec{k}}^{\,\phantom{\dagger}}}

\newcommand{\inte}{\int_{-\infty}^{\infty} {\rm d}\epsilon}

\newcommand{\partabl}[2]{\frac{{\rm \partial} #1}{{\rm \partial} #2}}

\begin{document}
\draft

\twocolumn[\hsize\textwidth\columnwidth\hsize\csname @twocolumnfalse\endcsname


\title{A model of semimetallic behavior in strongly correlated
electron systems}
\author{Stefan Blawid}
\address{Max-Planck-Institut f\"ur Physik komplexer Systeme,
N\"othnitzer Str.\ 38, 01187 Dresden, Germany}
\date{\today}

\maketitle


\begin{abstract}
  Metals with values of the resistivity and the Hall coefficient much
  larger than typical ones, e.g., of sodium, are called semimetals.
  We suggest a model for semimetals which takes into account the
  strong Coulomb repulsion of the charge carriers, especially
  important in transition-metal and rare-earth compounds.  For that
  purpose we extend the Hubbard model by coupling one additional
  orbital per site via hybridization to the Hubbard orbitals. We
  calculate the spectral function, resistivity and Hall coefficient of
  the model using dynamical mean-field theory. Starting from the
  Mott-insulating state, we find a transition to a metal with
  increasing hybridization strength (``self-doping''). In the metallic
  regime near the transition line to the insulator the model shows
  semimetallic behavior. We compare the calculated temperature
  dependence of the resistivity and the Hall coefficient with the one
  found experimentally for $\rm Yb_4As_3$. The comparison demonstrates
  that the anomalies in the transport properties of $\rm Yb_4As_3$
  possibly can be assigned to Coulomb interaction effects of the
  charge carriers not captured by standard band structure
  calculations.
\end{abstract}



\vskip2pc]

\section{Introduction}
\label{sec:intro}

Conventional semimetals like the pentavalent elements As, Sb and Bi
can be well understood using the approximation of independent
electrons. Having an even number of electrons per primitive cell,
these solids indeed come very close to being insulators. They are not,
because the occupied band nearest to the Fermi energy overlaps slighty
with the unoccupied band lowest in energy. Therefore a few holes and a
few electrons in these two bands can contribute to transport leading
to a very small number of charge carriers $n$, as seen in the measured
Hall coefficient $R_{\rm H} = 1/n{\rm ec}$~of these materials.
However, the independent electron picture is often not appropiate to
describe transition-metal and rare-earth compounds. Prominent examples
are the transition-metal oxides like $\rm V_2O_3$, $\rm MnO$ or $\rm
NiO$. These oxides are insulators although they contain partially
filled $d$ bands. Already in 1949, Mott identifies the Coulomb
interaction of the conduction electrons as reason for the insulating
behavior \cite{Mott49}. Meanwhile, the term Mott insulator has become
a synonym for solids with localized charge carriers due to their
mutual Coulomb repulsion. In the early transition-metal oxides the
movement of the charge carriers is completely blocked due to their
mutual Colomb repulsion. It seems likely that there also exist
compounds in which the movement of the charge carriers is only nearly
blocked and which exhibit therefore not insulating but semimetallic
behavior. The aforementioned conventional semimetals As, Sb and Bi
are solids which come close to being band insulators. There might
exist transition-metal or rare-earth compounds which are semimetallic
because they come close to being a Mott insulator. In
Sec. \ref{sec:exp} we will discuss a possible example, the semimetal
$\rm Yb_4As_3$.

In this paper we want to investigate a simple model which is able to
mimic a semimetal coming close to being a Mott insulator. Therefore we
are heading for a model which can describe the transition from a Mott
insulator to a metal by varying some control parameter. The appealing
idea is that a metal near a quantum transition to an insulator behaves
like a semimetal. This is so, because the conductivity vanishes for an
insulator. Therefore resistivity and Hall coefficient should become
very large if one approach the insulating state from the metallic
side. It is important to note, that a non vanishing Hall coefficient
can be only obtained in a non particle-hole-symmetric model. This
excludes one possible candiate, the well-known Hubbard model with one
electron per site. Here the control parameter would be the ratio
between the kinetic and the Coulomb energy of the electrons which can
be changed by applying pressure. A suited model system for our purpose
is the self-doped Hubbard model introduced in
Ref. \onlinecite{Blawid96}.  Here additional orbitals are coupled to a
half-filled Hubbard model via hybridization. The additional electron
states serve as a generic reservoir of charge. The hybridization
drives the desired insulator-to-metal transition.

The self-doped Hubbard model is not based on purely academic
considerations. It is inspired by the experimental observation of
charge order in some transition-metal and rare-earth compounds like
$\rm Yb_4As_3$ and $\rm NaV_2O_5$. At high temperatures these
compounds are mixed-valent systems. In $\rm Yb_4As_3$ for example the
Yb-ions fluctuate between a twofold and threefold positive charged
state leading to a formal valency of $\rm Yb^{2.25+}$. At low
temperatures the charges avoid their mutual Coulomb repulsion and stay
as far apart as possible. In $\rm Yb_4As_3$ this leads to a static
charge ordered state in which the nearest-neighbor sites of an $\rm
Yb^{3+}$-ion are only occupied by $\rm Yb^{2+}$-ions and not by other
$\rm Yb^{3+}$-ions
\cite{Fulde95,Schmidt96a,Schmidt96b,Schmidt96c,Kohgi97,Rams96}.
Therefore, the $\rm Yb^{3+}$-ion are confined on a subsystem of all
possible sites, here chains in $\langle 111 \rangle$-direction. A
charge lattice is superposed on the crystal lattice.  In $\rm
Yb_4As_3$ the charge order leads to a structural phase transition at
$T_c \approx 295$K. The foremost equivalent Yb-sites split up in $\rm
Yb^{2+}$-sites with a crystal environment of lower symmetry and $\rm
Yb^{3+}$-site with one of higher symmetry. The superposed charge
lattice manifest itself in the distorted crystal structure below
295K. It is important to note, that charge ordering drive a compound
towards a Mott-insulating state because the confinement of the charges
to a sublattice reduces their kinetic energy. Transition-metal or
rare-earth compounds whith a charge order transition are therefore
promising candidates of systems coming close to being a Mott
insulator. Indeed, $\rm Yb_4As_3$ is a semimetal \cite{Ochiai90}. A
very simplified model have to contain sites preferential occupied by
the charges representing the sublattice superposed by charge
order. The charges are able to leave this sites to gain some kinetic
energy. Therefore, the sites lower in energy have to be coupled to an
empty reservoir of charges representing all other accessible but
predominantly unoccupied sites of the crystal. The self-doped Hubbard
model is a possible realization of this idea. To take advantage of
sophisticated many-body techniques, namely the dynamical mean-field
theory \cite{Pruschke95,Georges96}, the empty reservoir of charges in
the self-doped Hubbard model is simply represented by unoccupied
orbitals hybridizing purely local with the ones of the occupied sites.

The temperature behavior of the resistivity and the Hall coefficient
is anomalous in $\rm Yb_4As_3$. Both quantities show a non-monotonous
temperature dependence and exhibit a maximum at a characteristic
temperature. The characteristic temperatures differ for the two
properties. Therefore, not only the maximum value of the resistivity
and the Hall coefficient of the self-doped Hubbard model will be of
interest but also the temperature dependence of the two properties.
In this paper we calculate the density of states, resistivity and Hall
coefficient of the self-doped Hubbard model in the vicinity of the
metal-to-insulator transition using dynamical mean-field theory. In
the next section we introduce the model and discuss the mapping of
this multi-band model to an impurity model within the dynamical
mean-field theory. The impurity model is studied numerically by an
extension of the non-crossing approximation to a two-orbital
impurity. We present the results in Sec. \ref{sec:results}, discuss
the applicability of the self-doped Hubbard model to $\rm Yb_4As_3$ in
Sec. \ref{sec:exp} and finally conclude in Sec. \ref{sec:conc}.

\section{Model and Method}
\label{sec:model}

The self-doped Hubbard model is given by
\begin{eqnarray} 
\label{eq.smsk}
H & = & -(\Delta+\mu) \sum_{\vec{k},\sigma} \ffkd \ffk 
       + \sum_{\vec{k},\sigma} \dis \, \ffkd \ffk
       -\mu \sum_{\vec{k},\sigma} \cckd \cck \nonumber\\   
 & &   + V \sum_{\vec{k},\sigma} \left(\,\,\ffkd \cck + {\rm h.c.} \right)
       + U \sum_i \niup \nidown \; \; ,
\end{eqnarray}

The model consists of a lattice of $f$ orbitals described by the
Hubbard Hamiltonian with on-site Coulomb repulsion $U$. Locally the
$f$ orbitals hybridize with additional orbitals called $c$. The
hybridization strength is $V$. For an electron it is energetic
favourable to occupy an $f$ orbital. The energy difference between an
occupied $c$ orbital and an occupied $f$ orbital is given by the
charge-transfer energy $\Delta$. The filling is one electron per site,
i.e., the system is quarter-filled.  For $V=0$ the self-doped model
reduces to a half-filled one-band Hubbard model. The Coulomb repulsion
of the electrons is large, $U>U_{\rm c}$, and the system is a Mott
insulator.  Here we have assumed a finite critical value of the
Mott-Hubbard transition.  For $V \neq 0$ the mean occupation of $f$
orbitals is smaller than one. We may say that for $V=0$ the electrons
order in the sense that they only occupy the $f$ subsystem.  The
charge ordered state is an insulator. A finite hybridization hinders
the order to be perfect.

In Ref. \onlinecite{Blawid96} we have shown that for $V=\infty$ the
self-doped Hubbard model again reduces to a half-filled one. However,
the effective Coulomb repulsion is reduced to $U/4$ and the effective
hopping to $t/2$. In a regime of $U_{\rm c} < U < 2 U_{\rm c}$ we
therefore expect an insulator-to-metal transition to take place with
increasing value of $V$. Unfortunately simple approximation schemes
like slave-boson mean-field and alloy-analog approximation fail to 
reproduce this quantum transition. In this paper we apply the
dynamical mean-field theory to the model.

Though we are aiming at the resistivity and the Hall coefficient of
the model we first calculate the one-particle Green's
function $G(\vec{k},\omega)$. Given the noninteracting Green's function,
$G_0(\vec{k},\omega)$, of the self-doped Hubbard model
\begin{equation}
G_0^{-1}(\vec{k},\omega) = \left( 
\begin{array}{cc}
\omega+\mu+\Delta-\epsilon_{\vec{k}} & -V \\
-V & \omega+\mu
\end{array}
\right)
\end{equation} 
the self-energies are defined by Dyson's equation  
\begin{equation}
G(\vec{k},\omega) = \left( G_0^{-1}(\vec{k},\omega) - 
\Sigma(\vec{k},\omega) \right)^{-1}.
\end{equation}
The dynamical mean-field theory assumes a momentum-independent
self-energy $\Sigma(\vec{k},\omega) \rightarrow \Sigma(\omega)$. In
general the self-energy can be considered as a functional
\cite{Baym62} of the full Green's function $G(\vec{k},\omega)$. In
particular the local approximated self-energy is only a functional of
the local Green's function
\begin{equation}
\label{eq:local}
G(\omega) = \frac{1}{N} \sum_{\vec{k}} G(\epsilon_{\vec{k}},\omega).
\end{equation}
The functional dependence is generated purely by the interaction term
in the Hamiltonian. It is thus the same for an impurity model with the
same interactions. Therefore, lattice and impurity model have the same
self-energy provided we identify the Green's function ${\cal
  G}(\omega)$ of the impurity model with $G(\omega)$. In the case of
the self-doped Hubbard model the corresponding impurity model reads

\begin{table*}
\begin{tabular}{cl}
  \rule{0mm}{8mm} (i) & $G_{ff}(\omega) = \sum_{nm}
  |P_{\,1}^{\,\sigma}(n,m)|^2\, \greew{X_{nm}}{X_{mn}}$ \\[0.2cm] 
  (ii) & $\greew{X_{nm}}{X_{mn}} = \frac{1}{Z_{\rm Imp}}
  \inte\,e^{-\beta\epsilon}
  \{p_n(\epsilon)\,P_m^{\phantom{*}}(\epsilon+\omega)
  -P_n^{*}(\epsilon-\omega)\,p_m(\epsilon)\}$\\[0.2cm] 
  (iii) & $Z_{\rm\,Imp} = \sum_m
  \inte\,e^{-\beta\epsilon}\,p_m(\epsilon)$\\[0.2cm] 
  (iv) & $p_m(\epsilon) = -\frac{1}{\pi}\,{\rm Im}\,P_m(\epsilon+{\rm
    i}0^{+})\;,$
  $\;\;\;P_m(z)=\inte\,\frac{p_m(\epsilon)}{z-\epsilon}$ \\[0.2cm] 
  (v) & $P_n(z) = \frac{1}{z-E_n-\Sigma_n^{(2)}(z)}$ \\[0.4cm] 
  (vi) & $\Sigma_n^{(2)}(z) = -\frac{1}{\pi} \sum_{m\sigma} \inte\,{\rm Im}
  {\cal J}(\epsilon+{\rm i}0^{+}) \left\{ |P_{\,1}^{\,\sigma}(n,m)|^2
  f(\epsilon) P_m(z+\epsilon)+|P_{\,1}^{\,\sigma}(m,n)|^2 f(-\epsilon)
  P_m(z-\epsilon) \right\}$
\end{tabular}
\vspace*{0.5cm}
\caption{\label{tab:nca}Summary of the used equations of the
  generalized NCA to solve approximatively the impurity problem}
\end{table*}

\begin{eqnarray}
\label{eq:imp}
H_{\rm eff} & = & H_{\rm cell} + H_{\rm med}\\[0.2cm]
H_{\rm cell} & = & 
-(\Delta+\mu) \sum_{\sigma} \nsig + U \, \nup \ndown 
- \mu \sum_{\sigma} \ncsig \nonumber \\
 & & + V \sum_{\sigma} \left(\,\,\ffd \cc + {\rm h.c.} \right) \\ 
H_{\rm med} & = & 
\sum_{\vec{k}\sigma} E_{\vec{k}}^{\,\phantom{\dagger}} \, 
d_{\vec{k}\sigma}^{\,\dagger} d_{\vec{k}\sigma}^{\,\phantom{\dagger}}
+ \sum_{\vec{k}\sigma}
\left(\,\,\Gamma_{\vec{k}}^{\,\phantom{\dagger}} \, 
d_{\vec{k}\sigma}^{\,\dagger} \ff + {\rm h.c.} \right)\;.
\nonumber
\end{eqnarray}
Note, $H_{\rm eff}$ just embeds a single unit-cell of the
original lattice model in an effective medium. Following the
discussions in Ref. \onlinecite{Schork97} and \onlinecite{Schork98} it
is easy to show that the matrix equation
\begin{equation}
\label{eq:selfcon}
G(\omega)={\cal G}(\omega)
\end{equation}
can be fulfilled by choosing a single function
\begin{equation}
{\cal J}(\omega) = \sum_{\vec{k}} \frac{|\Gamma_{\vec{k}}|^2}
{\omega-E_{\vec{k}}}
\end{equation}
which describes the coupling of the $f$ orbital to the bath. The
integration in Eq. (\ref{eq:local}) can be performed analytically if
we choose a semielliptic density of states
\begin{equation}
\rho_0(\epsilon) = \frac{1}{N} \sum_{\vec{k}} \del{\epsilon-\dis}
=\frac{2}{\pi\,W^2} \,\sqrt{W^2-\epsilon^2}\; . 
\end{equation}
The self-consistency (\ref{eq:selfcon}) then reduces to
\begin{equation}
{\cal J}(\omega) =  \frac{W^2}{4^{\phantom{2}}} \, G_{ff}(\omega) \; .
\end{equation}
In all our calculation we choose $W=1$ as unit of energy.  

What remains is the calculation of the one-particle Green's function
of the impurity problem (\ref{eq:imp}).  Here we make use of an
extension of the non-crossing approximation \cite{Bickers87a} (NCA) to
the case of more than two ionic propagators.  Note, that the
``impurity'' in (\ref{eq:imp}) has 16 eigenstates
\begin{equation}
\label{eq:dia}
H_{\rm cell} = \sum_{m=1}^{16} E_m \, X_{mm}
\end{equation}
and not only two as in the case of the $U=\infty$ impurity Anderson
model. In Eq. (\ref{eq:dia}) we have introduced the Hubbard operators
$X_{nm} = \hubb{n}{m}$. The generalized NCA has been applied
successfully to the finite-$U$ impurity Anderson model
\cite{Pruschke89}, the Emery model in the dynamical mean-field theory
\cite{Lombardo96}, and the Anderson-Hubbard \cite{Schork97} as well as
the Kondo-Hubbard model \cite{Schork98} in this approximation. We just
summarize the basic equations in Tab. \ref{tab:nca} and refer the
reader for further details to the literature, e.g. see Refs.
\onlinecite{Lombardo96,Schork97,Schork98}. The coupled integral
equations (v) and (vi) in Tab. \ref{tab:nca} are solved numerically by
introducing defect propagators \cite{Bickers87b} and making use of the
fast Fourier transformation \cite{Lombardo96}. The $f$ Green's
function of the impurity model gives a new function ${\cal J}(\omega)$
via Eq. (\ref{eq:selfcon}) and the calculations are iterated until
self-consistency is achieved.

We are now turning to the calculation of the resistivity and the Hall
coefficient. In the self-doped Hubbard model the electrons can hop
only from one $f$ orbital on site $i$ to an $f$ orbital on a
neighboring site $i+\delta_\alpha$ ($\alpha = x,y,z$ denotes the
direction in space). The current operator of the model is therefore
given by
\begin{equation}
\label{eq:current}
  j_{i+\frac{\delta_{\alpha}}{2}}^{\,\alpha} =  \frac{{\rm
    i}\,}{{\rm \hbar}}\,a\,t\, \sum_{\sigma}
  (f_{i+\delta_{\alpha}\,\sigma}^{\,\dagger} \ffi - \ffid
  f_{i+\delta_{\alpha}\,\sigma}^{\,\phantom{\dagger}})\;, 
\end{equation}
where $a$ is the lattice spacing. In momentum space the current
operator reads 
\begin{equation}
\label{eq:currentq}
  j_{\vec{q}}^{\,\alpha} =  \frac{1}{{\rm \hbar}} \sum_{\vec{p}\sigma}
  \left( \partabl{\epsilon_{\vec{p}}}{p^{\alpha}}\right)
    f_{\vec{p}-\frac{\vec{q}}{2}\,\sigma}^{\,\dagger}
    f_{\vec{p}+\frac{\vec{q}}{2}\,\sigma}^{\,\phantom{\dagger}}\;.
\end{equation}

Given the current operator we can straightforward repeat the
calculation of the conductivity in the case of the Hubbard model. It
is not surprising that we end with the same expressions as in the case
of the Hubbard model \cite{Pruschke95} just inserting the $f$ spectral
density
\begin{eqnarray}
\label{eq:leitxx}
\sigma^{xx} & = & \frac{{\rm e^2} \pi}{6 {\rm \hbar} a}
\int_{-\infty}^{\infty} {\rm d}\omega \left(-\partabl{f}{\omega}\right)
\int_{-1}^{1} {\rm d}\epsilon\, \rho_{0}(\epsilon)\,
\rho_f^2(\epsilon,\omega) \\[0.2cm]
\label{eq:leitxy}
\sigma_{\rm H}^{xyz} & = & \frac{{\rm |e|^3} \pi^2 a}{27 {\rm \hbar}^2
  {\rm c}}
\int_{-\infty}^{\infty} {\rm d}\omega \left(-\partabl{f}{\omega}\right)
\int_{-1}^{1} {\rm d}\epsilon\, \rho_{0}(\epsilon)\,
\epsilon\, \rho_f^3(\epsilon,\omega)\;.
\end{eqnarray}
Here $f(\omega) = \left[ \exp(\beta \omega) + 1 \right]^{-1}$ is the
Fermi function. The derivation involves two steps. First, we consider
the noninteracting case like in Ref. \onlinecite{Voruganti92}. Second,
we mark that vertex corrections in the linear response diagrams
vanish.  Here the fact enters that the one-particle self-energy is
momentum independent. In addition the special $\vec{k}$-dependence of
the free propagators via $\epsilon_{\vec{k}}$ and of the vertices via
$\partabl{\epsilon_{\vec{k}}}{k^{\alpha}}$ (see Eq.
(\ref{eq:currentq})) is needed. \cite{Pruschke95} Because the vertex
corrections vanish the free propagators in the expression for the
noninteracting case can simply be replaced by full ones.

Given the conductivity $\sigma^{xx}$ and the Hall conductivity
$\sigma_{\rm H}^{xyz}$ we know the resistivity $\rho_{xx} =
1/\sigma^{xx}$ and the Hall coefficient $R_{\rm H}=\sigma_{\rm
  H}^{xyz}/(\sigma^{xx})^2$. The units of the two transport
properties are given by $[\rho_{xx}] \approx [\hbar a/{\rm e^2}]
\approx \rm 0.1 m\Omega cm$ and $[R_{\rm H}] \approx [a^3/{\rm e c}]
\approx \frac{1}{\rm c} 10^{-5} \rm cm^3/C$, respectively. Note, the
conductivities are given by the one-particle spectral function only
\begin{equation}
\rho_f(\epsilon,\omega) = -\frac{1}{\pi}\;{\rm Im}\,
G_{ff}(\epsilon,w+i0^+)\;.
\end{equation}
In particular, if $\rho_f(\epsilon,\mu)$ vanishes for all energies
$\epsilon$, i.e., $\rho_f(\mu)=0$, than also the conductivity vanishes
for low temperatures. In this case the system is an insulator.

\section{Results}
\label{sec:results}

When discussing the results we will frequently refer to experimental
findings for the semimetal $\rm Yb_4As_3$. In the next Section we will
discuss the applicability of the self-doped Hubbard model to this
rare-earth compound.

\begin{figure}
\centerline{\psfig{file=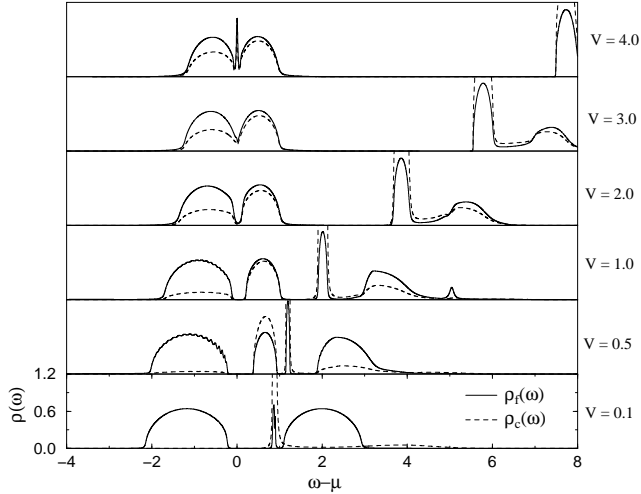,width=8cm,angle=270}}
\vspace*{1cm}
\caption{Spectral density of the $f$ and $c$ orbitals in the self-doped
  Hubbard model for $U=3$, $\Delta=2$ and different values of the
  hybridization $V$. For $V \geq 2.7$ the system is metallic
  and a sharp peak at the chemical potential is present}
\label{fig:spectra1}
\end{figure}
\begin{figure}
\centerline{\psfig{file=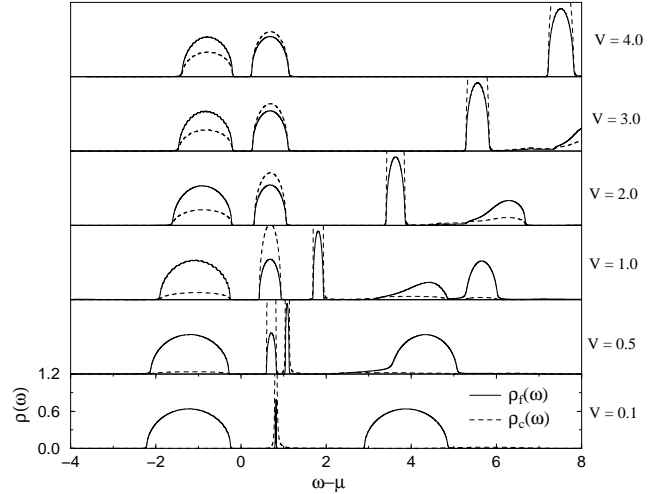,width=8cm,angle=270}}
\vspace*{1cm}
\caption{Spectral density of the $f$ and $c$ orbitals in the self-doped
  Hubbard model for $U=5$, $\Delta=2$ and different values of the
  hybridization $V$. The system stays insulating independent of the
  value of $V$.}
\label{fig:spectra2}
\end{figure}
\begin{figure}
\vspace*{-0.5cm}
\centerline{\psfig{file=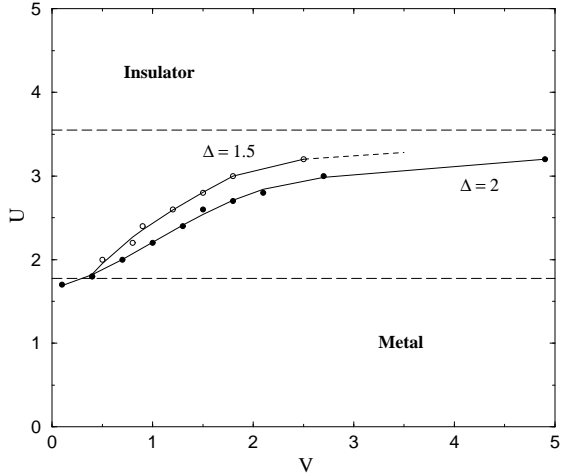,width=9cm,angle=270}}
\vspace*{0.5cm}
\caption{Metal-insulator diagram of the self-doped Hubbard model for
  $\Delta=2$ (full circles) and $\Delta=1.5$ (open circles). The lines
  are guide to the eye and separate the metallic from the insulating
  regime. Transitions between the two states are limited to a region
  between $U_c=1.775$ and $2 U_c = 3.55$ marked by horizontal dashed
  lines.}
\label{fig:phas}
\end{figure}

Heading for the metal-insulator diagram of the self-doped Hubbard
model we first concentrate on the value of the $f$ spectral function
at the chemical potential $\rho_f(\mu)$ for different values of the
hybridization.  In Fig. \ref{fig:spectra1} we display the evolution of
the spectral function of the model with increasing $V$ and fixed
values for $U$ and $\Delta$. The $f$ and $c$ spectral functions are
obtained by the treatment outlined in Sec.  \ref{sec:model}. Within
our approach the critical Coulomb repulsion of the Mott-Hubbard
transition in the one-band Hubbard model is $U_{\rm c} = 1.77(5)$. The
chosen value of $U=3$ fullfills $U_{\rm c} < U < 2 U_{\rm c}$. As
expected for this value of $U$ the model undergoes an
insulator-to-metal transition. For $V < 3$ we obtain $\rho_f(\mu)=0$
and the system is an insulator. For $V \geq 3$ we find a finite value
$\rho_f(\mu) \neq 0$ and the model is a metal. Increasing the
resolution in $V$ we obtain the value $V_{\rm c} \approx 2.7$ for the
critical hybridization of the transition. Next, we consider the case
of a very large Coulomb repulsion $U=5$, i.e., $U>2U_{\rm c}$. In Fig.
\ref{fig:spectra2} we demonstrate that the gap between the filled band
and the lowest empty band never closes. The density of states at the
chemical potential vanishes for all values of the hybridization.  This
shows clearly that the insulator-to-metal transition is restricted to
a parameter region $U_{\rm c} < U < 2 U_{\rm c}$. In Fig.
\ref{fig:phas} we display the resulting metal-insulator diagram of the
self-doped Hubbard model in the $(U,V)$-plane for two different values
of the charge transfer energy.

Qualitatively a phase diagram of this form was already predicted in
Ref. \onlinecite{Blawid96}. Indeed, the bands in the spectral
function of the lattice model evolve like the transition energies of
the single two-orbital impurity ($H_{\rm cell}$ in Eq.
(\ref{eq:imp})). Therefore no sophisticated calculations are required
to determine roughly the position of the centers of bands. We denote
the possible transitions of the single two-orbital impurity by
$(f^1c^0) \rightarrow (f^nc^m)$.  The correspondence of the respective
transition energies and the band energies allows us to classify the
occupied band below the chemical potential as $(f^0c^0)$-band and the
unoccupied bands above $\mu$ as $(f^1c^1)_{\rm s}$-, $(f^1c^1)_{\rm
  t}$- and $(f^2c^0)$-bands (here s and t refers to singlet and
triplet).  Lattice effects show up in the ratio between $f$ and $c$
weight constituting a single band. For example, the occupation number
of $c$ orbitals is larger in the lattice model than in the two-orbital
impurity marking a larger $c$ weight in the occupied $(f^0c^0)$-band.
This is a simple consequence of quantum fluctuations between different
impurity configurations in the ground state of the lattice model.

\begin{figure}
\centerline{\psfig{file=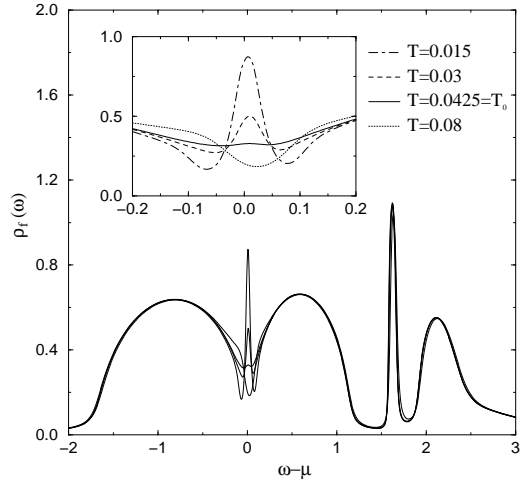,width=8cm,angle=270}}
\vspace*{1cm}
\caption{Temperature dependence of the $f$ spectral density of the
  self-doped Hubbard model with parameters $(U,\Delta,V) =
  (1.8,2,0.6)$. The inset shows the enlarged frequency region close to
  the chemical potential. Below the temperature $T_0$ a sharp peak
  arises at the chemical potential.}
\label{fig:temp}
\end{figure} 
Nevertheless, the most prominent feature of the lattice model is the
appearance of a sharp Kondo-like resonance close to the chemical
potential.  The temperature dependence of the resonance is shown in
\mbox{Fig. \ref{fig:temp}}. The peak arises below a characteristic
temperature $T_0$. Regarding the hybridization dependence of the
resonance we state that it becomes narrower and shows up at a
lower temperature $T_0(V)$ when we approach the insulating state from
the metallic side, i.e., when we regard the limit $V \rightarrow
V_{\rm c}^+$. Note, that for temperatures just above $T_0$
$\rho_f(\mu)$ is a monotonously decreasing function in this limit.
Therefore the observed behavior of $T_0$ is very similar to the
well-known dependence of the Kondo temperature on the density of
states $T_{\rm K} \sim \exp{\left(-b/\rho(\mu)\right)}$. The usual
Kondo effect can be interpreted in terms of the resonant-level
model.\cite{Hewson} The conduction electrons with energies close to
$\mu$ are virtually bound to localized electrons due to resonance
scattering. In a similar way one may assign the resonance seen in our
calculations to a band-Kondo effect where the conduction electrons are
virtually bound not to localized but to moving electrons. This
``band-Kondo effect'' is a typical feature of strongly correlated
metals treated within the dynamical mean-field theory. It occurs also
in the one-band Hubbard model. It is still a matter of dispute if the
effect is physical or a shortcoming of the mapping of the lattice
model to an impurity model as done within the dynamical mean-field
theory. But at least in the case of the Hubbard model there are
indications that the effect can be seen also by other methods.
\cite{Bulut94,Preuss95}
 
\begin{figure}
\centerline{\psfig{file=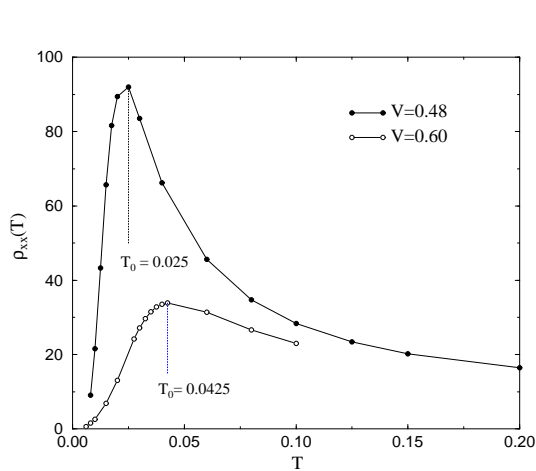,width=8cm,angle=270}}
\vspace*{1cm}
\caption{Temperature dependence of the specific resistivity in the
  self-doped Hubbard model. The unit of the the resistivity is $\rm
  0.1 m\Omega cm$. Magnitude and position of the maximum shift with
  the value of the hybridization $V$.}
\label{fig:res}
\end{figure}
The strong temperature dependence of the spectral density close to the
chemical potential causes a strong temperature dependence of the
resistivity. In Fig. \ref{fig:res} we show this dependence as obtained
by the treatment outlined in Sec. \ref{sec:model}. First we focus on
the magnitude of the resistivity. The resistivity increases when we
decrease the hybridization, i.e., when we approach the insulating
regime. In fact we expect a diverging resistivity in the limit $V
\rightarrow V_{\rm c}^+$ because the conductivity vanishes in an
insulator. This is the key idea. In the vicinity of a quantum
transition to an insulator we can obtain large values for the
resistivity and the Hall coefficient implying semimetallic behavior.
In Fig. \ref{fig:res} the order of magnitude of the resistivity is
$\rm 1 m\Omega cm$, i.e., the same as observed in $\rm Yb_4As_3$. Note,
that large values of $\rho$ and $R_{\rm H}$ do not imply that the
occupation number of the $c$ orbitals $n^c=1-n^f$ have to be small.
Large values are possible for every $n_c$ in striking contrast to a
simple Drude picture when assuming the charge carriers in the system
are given by the missing electrons in the $f$ subsystem. We stress
this point because in $\rm Yb_4As_3$ there is experimental evidence
\cite{Kohgi97} that the number of missing holes in the $\langle 111
\rangle$-chains is not identical with the low carrier concentration
obtained from the large Hall coefficient.

As function of temperature the resistivity shows a maximum at a
characteristic temperature $T_0$. This temperature is identical with
the one discussed above, i.e., the temperature where the sharp
resonance arises in the spectral function. As seen in Fig.
\ref{fig:res} and as discussed before $T_0$ is larger for parameter
values deeper in the metallic regime of the self-doped Hubbard system.
$\rm Yb_4As_3$ can be tuned to a more insulating or metallic state
(indicated by the value of the resistivity) by applying pressure
\cite{Okunuki95} or substituting P or Sb for As.
\cite{Ochiai97,Aoki97} Indeed, the experimentally observed
characteristic temperature shows the expected behavior. In samples
with a lower (more metallic) resistivity $T_0$ becomes larger as
compared to samples with a higher resistivity. Note, that the
resonance appears also in the $c$ spectral function (see Fig.
\ref{fig:spectra1} or Fig. \ref{fig:reso}). Concerning $\rm Yb_4As_3$
we may interpret the c orbitals as As $p$ band with zero
bandwidth. However, in $\rm Yb_4As_3$ the $p$ band is broad and
especially the holes in this band should contribute to transport
\cite{Antonov98}. We may conjecture that also the $p$ holes reveal the
described low temperature scale $T_0$ (see also the discussion in the
next Section). This is known for the case of a periodic Anderson model
where also a non-monotonous temperature dependence of the resistivity
is obtained. \cite{Czycholl93}

The enhanced $f$ spectral function close to the chemical potential
indicates heavy masses of the charge carriers at low
temperatures. Unfortunately, our approach does not allow us to perform
the limit $T \rightarrow 0$. It is well-known \cite{Cox88} that the
non-crossing approximation underestimates the absolute value of the
imaginary part of the self-energy of the single-impurity Anderson
model. This failure will be enhanced in the self-consistent adjustment
of the hosting bath of conduction electrons leading to an unphysical
change of sign of the self-energy at low temperatures.
\cite{Pruschke93} Therefore we cannot proof the relation $\rho(T) = A
T^2$ and we cannot calculate the coefficient $A$. However, note that
in Fig.  \ref{fig:res} the onset of a $T^2$-behavior can be seen at
least for the case of $V=0.6$.

\begin{figure}
\centerline{\psfig{file=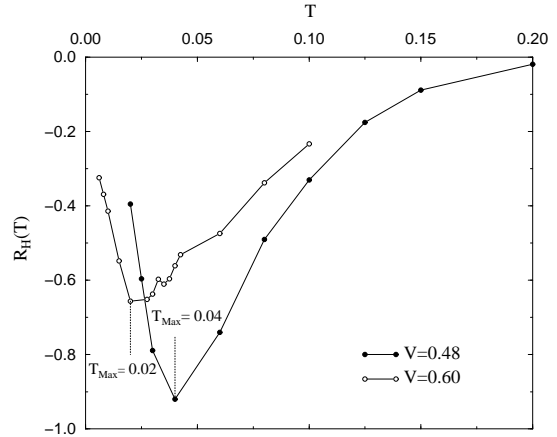,width=8cm,angle=270}}
\vspace*{1cm}
\caption{Temperature dependence of the Hall coefficient in the
  self-doped Hubbard model. The unit of the Hall coefficient is
  $\frac{1}{\rm c} 10^{-5} {\rm cm^3/C}$. Magnitude and position of
  the minimum shift with the value of the hybridization $V$.}
\label{fig:hall}
\end{figure}
We now turn to the temperature dependence of the Hall coefficient as
displayed in \mbox{Fig. \ref{fig:hall}}. For discussion we want to
compare qualitatively the calculated Hall coefficient with the one
measured in $\rm Yb_4As_3$.  First of all the Hall coefficient of the
self-doped Hubbard model is negative whereas it is positive in $\rm
Yb_4As_3$.  From our point of view this has to be expected. In the
model electrons order in the sense that they prefer to occupy the $f$
subsystem. In $\rm Yb_4As_3$ however holes in the $4f$-shells of the
Yb-ions are predominantly confined to the $\langle 111
\rangle$-chains. So there is a different sign for the charge carriers
which should show up in the Hall coefficient (see also the modified
model in the next Section). Second, the magnitude of the Hall
coefficient seems to be wrong. In Fig.  \ref{fig:hall} it is still
five orders of magnitude too small. This rather small value of the
Hall coefficient is caused by the still very symmetric spectral
function close to the chemical potential (see Fig. \ref{fig:temp}). In
this respect the chosen values of the hybridization (about $25\%$ of
the unperturbed $f$ bandwidth $2W$) are too large.  Nevertheless, as
we have stressed before the value of the Hall coefficient can be tuned
by choosing parameter values of the model closer to the
metal-to-insulator transition. Close to the transition arbitrary large
values for the Hall coefficient may be obtained.

Keeping this in mind we find indeed an anomalous temperature
dependence of the Hall coefficient as in experiment with a
characteristic temperature $T_{\rm Max}$. In the case of $V=0.6$ in
Fig. \ref{fig:hall} this characteristic temperature is smaller than
the one found in the resistivity (see Fig. \ref{fig:res}). The
relation $T_0/T_{\rm Max} \approx 2.13$ is close to the experimental
one $T_0/T_{\rm Max} \approx 1.75$. As we will point out in the next
Section, the self-doped Hubbard model is not a model simplifying the
band structure of $\rm Yb_4As_3$ close to the chemical
potential. Therefore, we only want to discuss possible implications of
the Coulomb interaction of the charge carriers on the transport
properties. If we would e.g.~identify the temperature $T_0 = 0.0425$
with the experimental one of 140K we obtain for the $f$ bandwidth $2W
\approx 0.6 \rm eV$. The LSDA+$U$ approach which will be reviewed in
the next Section gives a value of $2W=0.007 \rm eV$. But even
qualitatively there is still an essential difference.  Surprisingly
$T_{\rm Max}$ decreases with increasing value of the hybridization.
This behavior of $T_{\rm Max}$ is just opposite to the one of $T_0$.
In experiment both $T_0$ and $T_{\rm Max}$ behave the same. Especially
the ratio $\rho(T)/R_{\rm H}(T)$, i.e., the inverse Hall mobility of
the charge carriers, is independent of the applied pressure. It is not
clear if our finding for the behavior of $T_{\rm Max}$ in the model
depends on the parameter regime and possibly change closer to the
metal-to-insulator transition or not.

\section{The semimetal \protect{$\rm Yb_4As_3$}}
\label{sec:exp}

In the preceding Section we have presented the semimetal $\rm
Yb_4As_3$ as possible candidate for a compound which can be described
by the self-doped Hubbard model. In this Section we will describe the
properties of this compound in more detail. We want to argue that it
is indeed justified to compare the transport properties of the
self-doped Hubbard model and of $\rm Yb_4As_3$ as we have already
done.

$\rm Yb_4As_3$ is an example of a low-carrier heavy-fermion system
\cite{Ochiai90}. Below 100K a linear specific heat is found with a
large coefficient $\gamma \approx 200\,{\rm mJ/(molK^2)}$. The
Sommerfeld-Wilson ratio is of order unity. The large Hall coefficient
at low temperatures indicates an extremely small concentration of
$\delta = 0.3 \times 10^{-3}$ positive charge carriers/Yb-atom. The
large Hall coefficient and the large residual resistivity of $1\,{\rm
m\Omega cm}$ at low temperatures classify $\rm Yb_4As_3$ as a
semimetal. In $\rm Yb_4As_3$, all $\rm Yb$-atoms are aligned on four
families of chains. Assuming trivalent As the ratio of Yb-ions is $\rm
Yb^{3+}:Yb^{2+} = 1:3$. A $\rm Yb^{3+}$-ion has one hole in the $f$
shell ($4f^{13}$) and shows a magnetic moment due to Hund's rule
coupling. In a sequence of papers
\cite{Fulde95,Schmidt96a,Schmidt96b,Schmidt96c} Schmidt, Thalmeier and
Fulde suggests that at low temperatures the $\rm Yb^{3+}$-ions are
confined to one of the four chain systems, e.g., parallel $\langle 111
\rangle$-chains. The low energy excitations of these
quasi-one-dimensional spin chains explain the specific heat including
the large $\gamma$-coefficient.  However, due to the strong Coulomb
repulsion of the holes and the large distance of the $\rm
Yb^{3+}$-sites within one chain a perfect ordered state would imply an
insulator. Because $\rm Yb_4As_3$ is a semimetal a small fraction of
holes have to be redistributed either on the As-atoms or on the other
three chain systems, i.e., the $\langle 111 \rangle$-chains have to be
``self-doped''. It is tempting to identify this small fraction of
holes with the low concentration of charge carriers seen in the Hall
coefficient. The charge ordering of the $\rm Yb^{3+}$-ions is
meanwhile experimentally confirmed. \cite{Kohgi97,Rams96} However, the
number of missing holes in the $\langle 111 \rangle$-chains cannot be
exactly measured. Note, that the polarized neutron diffraction
experiment suggest that this number is not small but of the order of
several percent.

During the completion of our work, the electronic structure of $\rm
Yb_4As_3$ has been investigated using energy band calculations within
the so-called LSDA+$U$ approach \cite{Antonov98}.  The calculations
take into account the possibility of hole occupation on the Yb-sites
of the $\langle 111 \rangle$-chains and on the As-sites but not on the
remaining three chains system. In result a very narrow $\rm Yb^{3+}$
$4f$ hole band is obtained close to the top of a broad As $p$ band.
The Fermi energy is pinned to the bottom of the $4f$ hole band. From
this band structure indeed a very small occupation number of holes on
the As-sites results comparable to the small carrier number estimated
from the Hall coefficient. This finding is independent from the chosen
value of $U$ over a wide range of several eV. One should mention that
the possible hole distributions in the used approach are too
restrictive to deal with valence fluctuating systems like $\rm
Yb_4Bi_3$ and $\rm Yb_4Sb_3$ which do not show a charge ordering
transition. Only the $p$ holes should contribute to transport
properties because there is no direct $ff$-hopping. This explain
correctly the sign and magnitude of the Hall coefficient.

Some questions remain intriguing. In agreement with Fermi-liquid
behavior the resistivity is found to be of the form $\rho(T) = \rho_0
+ AT^2$ with a large coefficient $A \approx 0.84 \mu\Omega{\rm
cm/K^2}$.  It is hard to understand which scattering mechanism of the
$p$ holes can lead to such high effective masses in transport. Not
understood at all is the temperature dependence of the resistivity and
the Hall coefficient of $\rm Yb_4As_3$. Both quantities show a
non-monotonous temperature dependence and exhibit a maximum at a
characteristic temperature. The characteristic temperatures differ for
the two properties.

\begin{figure}
\centerline{\psfig{file=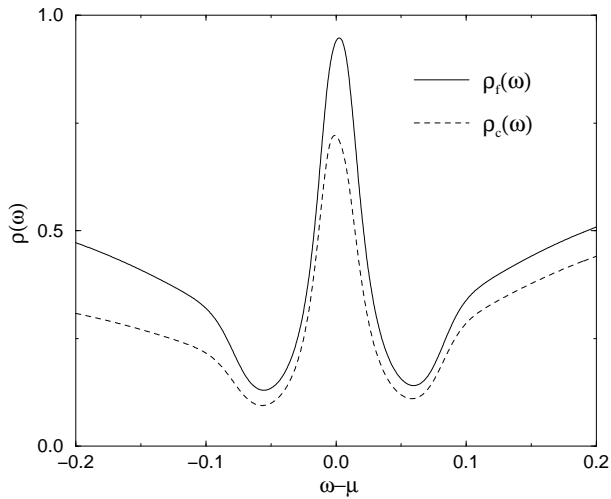,width=8cm,angle=270}}
\vspace*{0.5cm}
\caption{Spectral density of the $f$ and $c$ orbitals in the
  self-doped Hubbard model for $U=3$, $\Delta=2$ and $V=4$. In
  comparison with Fig. \protect{\ref{fig:spectra1}} the enlarged
  frequency region close to the chemical potential is shown.}
\label{fig:reso}
\end{figure}
Two comments seem to be appropiate. First of all, the band-structure
results question strongly the validity of the self-doped Hubbard model
for the compound under consideration $\rm Yb_4As_3$. In this compound
the transport seems to be purely due to the few charge carriers in the
nearly empty charge reservoir, e.g., the holes on the As-sites. In the
self-doped Hubbard model however the electrons in the $c$-states do
not contribute to the transport properties of the model at
all. Moreover, the perfect charge ordered state, e.g., all holes
confined to the $\langle 111 \rangle$-chains, is a Mott insulator of a
very extreme type because the kinetic energy of the holes on the
Yb-sites is nearly zero. Therefore, a suitable model Hamiltonian is
still of the form of Eq. (\ref{eq.smsk}) but with two important
changes. The filling should be three electrons per site and the
Coulomb repulsion should be suffered by the $c$-electrons. In this
modified version the $f$-subsystem (and not $c$) represents the As
sites with nearly no holes and the $c$-subsystem represents the $\rm
Yb^{3+}$-ions occupied by nearly one hole per site. In a future work
we will investigate this model. Nevertheless, we believe that the
modified model behaves quite similar to the self-doped Hubbard model.
Following the arguments outlined in Ref. \onlinecite{Blawid96} an
insulator-to-metal transition as function of the hybridization have to
be expected for both, the self-doped Hubbard model and the modified
model. In Fig. \ref{fig:reso} we show the $f$ and $c$ spectral
functions of the self-doped Hubbard model in the metallic regime. A
sharp peak close to the chemical potential is seen in both, the $f$
and the $c$ spectral function. The interaction induced many-body
effects of the correlated subsystem are carried over to the system of
the uncorrelated orbitals. Therefore, the similarity of the self-doped
Hubbard model with the proper model of $\rm Yb_4As_3$ holds despite
the fact that the conductivity of the modified model is given by the
spectral function of the uncorrelated orbitals. In conclusion, it is
justified to compare here some findings for the self-doped Hubbard
model with transport measurements of $\rm Yb_4As_3$.

We can argue in a slightly different way. In the charge ordered phase
of $\rm Yb_4As_3$ the Coulomb repulsion constrains the holes in the
$4f$-shell of the Yb-ions to be localised. The long range part of the
repulsion restricts the holes to chains in $\langle
111\rangle$-direction. The on-site part hinders the holes to move
along a single chain. This implies that the $4f$-spectral density is
split into bands separated by gaps which we may call charge order and
Mott-Hubbard gaps, respectively. The charge order gap will be of the
order of the transition temperature $T_c \approx 295$K and is much
smaller than the Mott-Hubbard gap of several eV. Therefore, the
relevant gap for electron excitations is the charge order gap and not
the Mott-Hubbard gap. However, the charge order gap in the
$4f$-spectral density is a true many-particle effect and cannot be
explained by the change of the unit cell between the charge disorderd
and the charge ordered phase. Note, the size of the unit cell is
unchanged in the charge order transition. The cell is only distorted,
changing from cubic to trigonal. The distortion of the unit cell has
only little influence on the electronic structure of $\rm Yb_4As_3$,
as shown in Ref. \onlinecite{Antonov98}. We believe that the charge
order gap resembles much more a Mott-Hubbard gap than it resembles a
gap in an usual band insulator.  The physics of the charge ordering is
not included in the self-doped Hubbard model.  But what is considered
in the used model is the effect of hybridization between a correlated
(Hubbard like) band and an uncorrelated band.

The second comment concerns the temperature dependence of the
resistivity and the Hall coefficient of $\rm Yb_4As_3$. On the first
sight it resembles the one found in other heavy-fermion compounds like
$\rm UPt_3$. In $\rm UPt_3$ the anomalous Hall conductivity arises
from skew-scattering of the charge carriers \cite{Kontani94}. As a
consequence, the anomalous Hall coefficient should be proportional to
the square of the resistivity for temperatures smaller than $T_{\rm
Max}$, the temperature for which a maximum in the Hall coefficient is
observed. However, in $\rm Yb_4As_3$ the relation $R_{\rm H} \sim
\rho^2$ is not fullfilled in the temperature regime from 80K down to
4K. Moreover, the zero-temperature value of the Hall coefficient in
$\rm Yb_4As_3$ is at least three orders of magnitude larger than in
other heavy-fermion compounds with a positive Hall coefficient. We
therefore exclude skew-scattering as possible mechanism to explain the
experimetal data.

\section{Conclusion}
\label{sec:conc}

In conclusion, we extended the Hubbard model by coupling one
additional orbital per site via hybridization to the Hubbard orbitals.
The Coulomb repulsion is supposed to be large, $U>U_{\rm c}$, where
$U_{\rm c}$ denotes the critical Coulomb repulsion for the
Mott-Hubbard transition in the one-band Hubbard model. We calculated
the spectral function, resistivity and Hall coefficient of the
``self-doped'' Hubbard model using dynamical mean-field theory. To
this end the lattice model is mapped onto an impurity model in which a
unit cell of the lattice is embedded self-consistently in a bath of
free electrons. The impurity model is studied numerically by an
extension of the non-crossing approximation to a two-orbital impurity.

The self-doped Hubbard model is an insulator only in a restricted
parameter regime. The hybridization with the added orbitals drives an
insulator-to-metal transition, provided the Coulomb repulsion
fullfills the constraint $U_{\rm c} < U < 2 U_{\rm c}$. This is a
correlation driven metal-insulator transition in a non particle-hole
symmetric case. In the vicinity of the transition to an insulator the
resistivity and the Hall coefficient become very large. Therefore the
model serves as a model for semimetallic behavior in systems where
the Coulomb interactions of the charge carriers nearly block their
movement. The number of missing electrons in the Hubbard subsystem,
i.e., the occupation number of the additional orbitals, can not be
interpreted as number of charge carriers seen in the transport
properties. A simple Drude analysis fails for this type of mechanism
leading to semimetallic behavior.

In the semimetallic regime the resistivity and the Hall coefficient
show an unusual temperature dependence. Both, $\rho(T)$ and $|R_{\rm
  H}(T)|$, exhibit a maximum at a characteristic temperature. The
characteristic temperatures for the resistivity $T_0$ differs from the
one of the Hall coefficient $T_{\rm Max}$. Depending on the parameter
regime either $T_0 < T_{\rm Max}$ or $T_0 > T_{\rm Max}$ is possible.
The characteristic temperatures change considerably when varying the
hybridization and tuning the system into a more insulating or metallic
state. At $T_0$ a resonance arises in the spectral function close to
the chemical potential. This resonance can be assigned to a Kondo-like
effect for band electrons. The enhanced spectral weight at the
chemical potential indicates heavy masses for the charge carriers.

We compared our findings with measurements of the resistivity and the
Hall coefficient for the semimetal $\rm Yb_4As_3$ including pressure
and substitution experiments. Although the self-doped Hubbard model is
not a simplified model extracted from band structure calculations we
found surprising similarities. We take this as indication that the
anomalous transport properties of $\rm Yb_4As_3$ can be indeed
assigned to the Coulomb interaction of the charge carriers which is
not treated adequate in standard band structure calculations. This
should be true even when the transport properties are dominated by As
$p$ holes. At least in the model the interaction induced many-body
effects of the Hubbard subsystem are carried over to the system of the
additional orbitals. Nevertheless, to obtain final conclusive results
the band structure of $\rm Yb_4As_3$ have to be used as input for a
simple model which still can be treated within the dynamical
mean-field theory. This is left for future work.

\acknowledgements

We would like to thank P.~Fulde, B.~Schmidt and P.~Thalmeier for
useful discussions and J.~Schmalian for numerical advice.


\end{document}